\documentclass[twocolumn,aps,pre,showpacs,eqsecnum,superscriptaddress,floatfix,twoside]{revtex4}

\usepackage{graphicx}
\usepackage{graphics}
\usepackage{amsmath}
\usepackage{amssymb}
\usepackage{mathrsfs}
\usepackage{epsf}

\begin{document}

\title{Dynamics of a metastable state
nonlinearly coupled to a heat bath driven by an external noise}

\author{Jyotipratim Ray Chaudhuri}
\email{jprc_8@yahoo.com} \affiliation{Department of Physics, Katwa
College, Katwa, Burdwan 713130, West Bengal, India}
\author{Debashis Barik}
\email{pcdb5@mahendra.iacs.res.in} \affiliation{Indian Association
for the Cultivation of Science, Jadavpur, Kolkata 700032, India}
\author{Suman Kumar Banik}
\email{skbanik@phys.vt.edu} \affiliation{Department of Physics,
Virginia Polytechnic Institute and State University, Blacksburg, VA
24061-0435}

\date{\today}

\begin{abstract}
Based on a system-reservoir model, where the system is nonlinearly
coupled to a heat bath and the heat bath is modulated by an external
stationary Gaussian noise, we derive the generalized Langevin
equation with space dependent friction and multiplicative noise and
construct the corresponding Fokker-Planck equation, valid for short
correlation time,  with space dependent diffusion coefficient to
study the escape rate from a metastable state in the moderate to
large damping regime. By considering the dynamics in a model cubic
potential we analyze the result numerically which are in good
agreement with the theoretical prediction. It has been shown
numerically that the enhancement of rate is possible by properly
tuning the correlation time of the external noise.
\end{abstract}


\maketitle


\section{Introduction}
Barrier crossing phenomena are ubiquitous and are a central issue in
many areas of natural science \cite{expt1,expt2}. Since the reaction
coordinate describing the transition across the barrier typically
interact with a large number of microscopic degrees of freedom, a
useful theory to start with is the construction of a Hamiltonian
which incorporates the effects of a heat bath environment. The
classical treatment of thermally activated barrier crossing
description is naturally accounted for by the generalized Langevin
equation (may be nonlinear) and by the corresponding Fokker-Planck
equation. Following Kramers \cite{kra,han1} transition rate can then
be calculated from a nonequilibrium steady state solution of the
Fokker-Planck equation describing a constant flux across the
potential barrier. Over several decades the Kramers' theory and many
of its variants has served as standard paradigms in various problems
of physical and chemical kinetics to understand the rate in
multidimensional systems in the overdamped and underdamped limits
\cite{lan,tak}, effects of anharmonicities \cite{van}, rate
enhancement by parametric fluctuations \cite{han2}, the role of
non-Gaussian white noise \cite{van,ski}, role of a relaxing bath
\cite{jrc1}, quantum and semiclassical corrections to classical rate
processes \cite{wol,jrc2,dsr} and related similar aspects.

The common feature of overwhelming majority of the aforesaid
treatments is that the system is thermodynamically closed, which
means that the noise of the medium is of internal origin so that the
dissipation and fluctuations gets balanced through the
fluctuation-dissipation relation. However, in a number of situations
the system is thermodynamically open, \textit{i.e.}, when the system
is driven by an external noise which is independent of system's
characteristic damping \cite{hor}. The main feature of the dynamics
in this case is the absence of any fluctuation-dissipation relation.
While in the former case a zero current steady state situation is
characterized by an equilibrium Boltzmann distribution, the
corresponding situation in the later case is defined only by a
steady state condition, if attainable.

A common approach to study the nonlinear, nonequilibrium systems
involves a description in terms of nonlinear stochastic differential
equation \cite{san1}. From a microscopic point of view, the
system-reservoir Hamiltonian description suggests that the coupling
of the system and the reservoir coordinates determines both the
noise and the dissipative terms in the Langevin equation describing
the dynamics of the system. If the system-bath interaction is linear
in the bath coordinates but arbitrary in the system coordinate, the
corresponding generalized Langevin equation incorporates a
multiplicative noise and consequently a nonlinear dissipative term
arises due to the nonlinear system-bath interaction \cite{lin}. A
canonical distribution for initial conditions of the bath variables
yields a zero average for the fluctuations and the
fluctuation-dissipation relation is again maintained \cite{lin}.
However, when the reservoir is modulated by an external noise, it is
likely that it induces fluctuations in the polarization of the
reservoir \cite{jrc3}. Since the fluctuations of the reservoir are
crucially dependent on the response function, one can envisage a
connection between the dissipation of the system and the response
function of the reservoir due to the external noise from a
microscopic stand point \cite{jrc3}. A direct driving of the system
usually breaks the fluctuation-dissipation relation and can generate a
biased directed motion that are seen in ratchet and molecular
motors \cite{aus}. On the other hand the bath modulation by an external
noise agency maintains a thermodynamic consistency relation, an analogue
of the fluctuation-dissipation relation of the closed system,
as a result of which the well known Kramers' turnover feature can be
restored \cite{jrc3}.

While the nonequilibrium, nonthermal systems have also been
investigated phenomenologically by a number of works in several
contexts \cite{jrc4}, these treatment are mainly concerned with
direct driving of the system by an external noise or a time
dependent field. In the present paper we consider a system-reservoir
model where the reservoir is modulated by an external noise and the
system is nonlinearly coupled to a heat bath thereby resulting a
nonlinear, multiplicative generalized Langevin equation. Our object here is to
explore the role of reservoir response on the system's dynamics and
to calculate the generalized escape rate from a metastable state for
a nonequilibrium open system in the presence of multiplicative
noise.

A number of different situations depicting the modulation of the
heat bath may be physically relevant. As for example, one may
experimentally study the reversible isomerization of
\emph{cis}-butene to \emph{trans}-butene, \emph{cis}-butene
$\rightleftharpoons$ \emph{trans}-butene. In terms of the reaction
rate theory both isomers represent the two stable local minima of
the potential energy landscape and are separated by the activation
energy which one stable configuration (\emph{cis} or \emph{trans})
needs to overcome to be converted into its isomeric form. To observe
the effect of external stochastic modulation one can carry out the
experiment in a photochemically active solvent (the heat bath) where
the solvent is under the influence of external monochromatic light
with fluctuating intensity of a wavelength which is absorbed solely
by the solvent molecules. As a result of it the modulated solvent
heats up due to the conversion of light energy into heat energy by
radiationless relaxation process and a effective temperature like
quantity develops due to the constant input of energy. Since the
fluctuations in the light intensity results in the polarization of
the solvent molecules, the effective reaction field around the
reactants gets modified. Provided the required stationarity of this
nonequilibrium open system is maintained the dynamics of the barrier
crossing becomes amenable to the present theoretical analysis that
follows.

Though the dynamics of a Brownian particle in a uniform solvent is
well-known, it is not very clear when the response of the solvent
becomes time dependent, as in a liquid crystal when projected onto
an anisotropic stochastic equation of motion or in the diffusion and
reaction in supercritical liquids and growth in living
polymerization \cite{res,her}. Also space-dependent friction is
realized from the presence of a stochastic potential in the Langevin
equation \cite{moi}. An exact Fokker-Planck equation for time and
space dependent friction was derived by Pollak \textit{et al}
\cite{pol}. Along with these formal developments, the theories of
multiplicative noise have found in wide applications in several
areas, \textit{e.g.}, activated rate processes, stochastic
resonance, laser and optics, noise induced transport etc \cite{mar}.
In passing we mention that the escape rate for a space dependent
friction is just not a theoretical issue but has been a subject of
experimental investigation over the last two decades \cite{har}.

The organization of the paper is as follows: In section II, starting
from a Hamiltonian description of a system nonlinearly coupled to a
harmonic reservoir which is modulated externally by a Gaussian
noise, we have derived the generalized Langevin equation with an
effective Gaussian noise $\xi(t)$. The statistical properties of
$\xi(t)$ has also been explored. In section III we have constructed
the corresponding Fokker-Planck equation, valid for small
correlation time, and have derived the generalized Kramers' rate for
moderate to large friction. In section IV, a specific example has
been carried out. Both the numerical and analytical results have
been analyzed in section V. The paper has been concluded in section
VI.


\section{The model: Heat bath modulated by external noise}

We consider a classical particle of mass $M$ nonlinearly coupled to
a heat bath consisting of $N$ harmonic oscillators driven by an
external noise. The total Hamiltonian of such a composite system can
be written as \cite{zwa,bra}

\begin{eqnarray}\label{2.1}
H & = & \frac{p^2}{2M}+V(x)+\frac{1}{2}\sum_{j=1}^{N}\left\{
\frac{p_j^2}{m_j}+m_j\omega_j^2\left(q_j-c_j g(x)\right)^2
\right\} \nonumber \\
& & +H_{int}
\end{eqnarray}

\noindent where $x$ and $p$ are the coordinate and the momentum of
the system particle, respectively, and $V(x)$ is the potential
energy of the system. $(q_j,p_j)$ are the variables for the $j$th
bath oscillator having frequency $\omega_j$ and mass $m_j$. $c_j$ is
the coupling constant for the system-bath interaction and $g(x)$ is
some analytic function of system coordinate. $H_{int}$ is the
interaction term between the heat bath and the external noise
$\epsilon (t)$, with the following form:

\begin{equation}\label{2.2}
H_{int}=\sum_{j=1}^{N} \kappa_j q_j \epsilon (t).
\end{equation}

\noindent The type of interaction we have considered between the
heat bath and the external noise, $H_{int}$ is commonly known as the
\textit{dipole interaction} \cite{dipole}. In equation (\ref{2.2}),
$\kappa_j$ denotes the strength of the interaction. We consider
$\epsilon(t)$ to be a stationary, Gaussian noise process with zero
mean and arbitrary correlation function

\begin{equation}\label{2.3}
\langle \epsilon(t) \rangle_e=0,\;\;\;\;\;\langle \epsilon (t)
\epsilon (t') \rangle_e = 2D\Psi (t-t')
\end{equation}

\noindent where $D$ is the external noise strength, $\Psi (t-t')$ is
the external noise kernel and $\langle...\rangle_e$ implies
averaging over the external noise processes.

Eliminating the bath degrees of freedom in the usual way \cite{lin}
(and setting $M$ and $m_j=1$) we obtain the generalized Langevin
equation

\begin{eqnarray}
\dot{x} &=& v \nonumber\\
\dot{v} &=& -\frac{d V(x)}{dx}-\frac{d g(x)}{dx} \int_0^{t} dt'
\gamma(t-t')\frac{d g(x(t'))}{dx(t')}v(t') \nonumber \\
        & & +\frac{dg(x)}{dx} \{ f(t)+\pi(t) \}\label{2.4}
\end{eqnarray}

\noindent where

\begin{equation}\label{2.5}
\gamma(t)=\sum_{j=1}^{N} c_j^2 \omega_j^2 \cos\omega_j t
\end{equation}

\noindent and $f(t)$ is the thermal fluctuation generated due to the
system-reservoir interaction and is given by

\begin{eqnarray}\label{2.6}
f(t) & = & \sum_{j=1}^{N}\left[ c_j \omega_j^2\left\{ q_j(0)-c_j
g(x(0)) \right\} \cos\omega_j t \right. \nonumber \\
     &   & \left. +\frac{v_j(0)}{\omega_j}\sin\omega_jt\right].
\end{eqnarray}

\noindent In equation (\ref{2.4}), $\pi(t)$ is the fluctuating force
generated due to the external stochastic driving $\epsilon(t)$ and
is given by

\begin{equation}\label{2.7}
\pi (t) = - \int_0^t dt' \varphi (t-t') \epsilon (t')
\end{equation}

\noindent where

\begin{equation}\label{2.8}
\varphi (t) = \sum_{j=1}^N c_j \omega_j \kappa_j \sin \omega_j t .
\end{equation}

\noindent
The form of equation (\ref{2.4}) indicates that the system is driven
by two forcing terms $f(t)$ and $\pi(t)$, both are multiplicative by
a function of system variable $dg(x)/dx$. Thus we have obtained a
generalized Langevin equation with multiplicative noise. To define
the statistical properties of $f(t)$, we assume that the initial
distribution is one in which the bath is equilibrated at $t=0$ in
the presence of the system but in the absence of the external noise
agency $\epsilon(t)$ such that

\begin{equation}\nonumber
\langle f(t) \rangle=0 \text{ and } \langle f(t) f(t') \rangle =k_B
T\gamma (t-t').
\end{equation}

\noindent Now at $t=0_{+}$, the external noise agency is switched on
and the bath is modulated by $\epsilon(t)$. The system dynamics is
governed by equation (\ref{2.4}), where apart from the internal
noise $f(t)$ another fluctuating force $\pi(t)$ appears that depends
on the external noise $\epsilon (t)$. So we define an effective
noise $\xi(t)[=f(t)+\pi(t)]$ whose correlation is given by \cite{jrc3}

\begin{eqnarray}\label{2.9}
\langle \langle \xi (t) \xi (t') \rangle \rangle & = & k_BT \gamma
(t-t') + 2D \int_0^t dt'' \int_0^{t'} dt'''
\nonumber \\
& & \times \varphi (t-t'') \varphi (t'-t''') \Psi (t''-t''')
\end{eqnarray}

\noindent along with $\langle \langle \xi (t) \rangle \rangle=0$,
where $\langle \langle ... \rangle\rangle$ means we have taken two
averages independently. It should be noted that the above equation
(\ref{2.9}) is not a fluctuation-dissipation relation due to the
appearance of the external noise intensity. Rather it serves as a
thermodynamic consistency condition. The statistical properties of
$\pi(t)$ are determined by the normal mode densities of the bath
frequencies, the coupling of the system with the bath, the coupling
of the bath with the external noise and on the statistical
properties of the external noise itself. Equation (\ref{2.7}) is
reminiscent of the familiar linear relation between the polarization
and external field, where $\pi(t)$ and $\epsilon(t)$ play the role
of the former and later, respectively. $\varphi(t)$ then can be
interpreted as a response function of the reservoir due to the
external noise $\epsilon(t)$. The structure of $\pi(t)$ suggests
that this forcing function, although obtained from an external
agency, is different from a direct driving force acting on the
system.

To obtain a finite result in the continuum limit, the coupling
function $c_i=c(\omega)$ and $\kappa_i=\kappa(\omega)$ are chosen
\cite{jrc3} as $c(\omega)=c_0/\omega\sqrt{\tau_c}$ and
$\kappa(\omega)=\kappa_0\omega\sqrt{\tau_c}$. Consequently
$\gamma(t)$ and $\varphi(t)$ reduces to the following forms:

\begin{equation}\label{2.10}
\gamma(t)=\frac{c_0^2}{\tau_c}\int d\omega \mathscr{D}(\omega)
\cos\omega t
\end{equation}

\noindent and

\begin{equation}\label{2.11}
\varphi(t)=c_0\kappa_0\int d\omega \mathscr{D}(\omega) \omega
\sin\omega t
\end{equation}

\noindent where $c_0$ and $\kappa_0$ are constants and $1/\tau_c$ is
the cutoff frequency of the oscillator. $\tau_c$ may be
characterized as the correlation time of the bath \cite{lin}. For
$\tau_c\rightarrow 0$ we obtain a delta-correlated noise process.
$\mathscr{D}(\omega)$ is the density of modes of the heat bath which
is assumed to be Lorentzian:

\begin{equation}\label{2.12}
\mathscr{D}(\omega)=\frac{2}{\pi}\frac{1}{\tau_c(\omega^2+\tau_c^{-2})}.
\end{equation}

\noindent
This assumption resembles broadly the behavior of the hydrodynamical
modes in a macroscopic system \cite{res}. With these forms of
$\mathscr{D(\omega)}$, $c(\omega)$ and $\kappa(\omega)$ we have the
expression for $\varphi(t)$ and $\gamma(t)$ as

\begin{subequations}
\begin{eqnarray}
\varphi(t)&=&\frac{c_0\kappa_0}{\tau_c}\exp(-t/\tau_c)\label{2.13a}\\
\gamma(t)&=&\frac{c_0^2}{\tau_c}\exp(-t/\tau_c)\label{2.13b}.
\end{eqnarray}
\end{subequations}

\noindent From equations (\ref{2.10}) and (\ref{2.11}) we obtain

\begin{equation}\label{2.14}
\frac{d
\gamma(t)}{dt}=-\frac{c_0}{\kappa_0}\frac{1}{\tau_c}\varphi(t) ,
\end{equation}

\noindent Equation (\ref{2.14}), an important content of the present
model, is independent of $\mathscr{D}(\omega)$. This expresses how
the dissipative kernel $\gamma(t)$ depends on the response function
$\varphi(t)$ of the medium due to the external noise $\epsilon(t)$.

If we assume that $\epsilon(t)$ is a $\delta$-correlated noise,
\textit{i.e.}, $\langle \epsilon(t)\epsilon(t')
\rangle_e=2D\delta(t-t')$ then the correlation function of $\pi(t)$
will be

\begin{equation}\label{2.15}
\langle \pi(t)\pi(t')
\rangle=\frac{Dc_0^2\kappa_0^2}{\tau_c}\exp(-|t-t'|/\tau_c)
\end{equation}

\noindent where we have neglected the transient terms
$(t,t'>\tau_c)$. This equation shows how the heat bath dresses the
external noise. Though the external noise is a $\delta$-correlated
noise, the system encounters it as an Ornstein-Uhlenbeck noise with
same correlation time of the heat bath but with an intensity
depending on the coupling $\kappa_0$ and the external noise strength
$D$. On the other hand, if the external noise is an
Ornstein-Uhlenbeck process with $\langle \epsilon(t)\epsilon(t')
\rangle_e=(D/\tau')\exp(-|t-t'|/\tau')$, the correlation function of
$\pi(t)$ is found to be

\begin{eqnarray}\label{2.16}
\langle \pi(t)\pi(t') \rangle & = &
\frac{Dc_0^2\kappa_0^2}{(\tau'/\tau_c)^2-1}\frac{\tau'}{\tau_c}\left\{
\frac{1}{\tau_c}\exp\left(-\frac{|t-t'|}{\tau'}\right) \right.
\nonumber \\
& & \left. -\frac{1}{\tau'}
\exp\left(-\frac{|t-t'|}{\tau_c}\right)\right\}
\end{eqnarray}

\noindent where we have neglected the transient terms. If the
external noise-correlation time be much larger than the internal
noise-correlation time, \textit{i.e.}, $\tau'\gg \tau_c$, which is
more realistic, then the dressed noise is dominated by the external
noise and we have from (\ref{2.16})

\begin{equation}\label{2.17}
\langle \pi(t)\pi(t')
\rangle=\frac{Dc_0^2\kappa_0^2}{\tau'}\exp\left(-\frac{|t-t'|}{\tau'}\right)
.
\end{equation}

\noindent On the other hand, when the external noise correlation
time is smaller than the internal one, we recover (\ref{2.15}).


\section{Generalized Fokker-Planck description and Kramers' escape rate}

To start with we consider the internal dissipation is Markovian
(\textit{i.e.} $\tau_c\rightarrow 0$ and the internal noise is
Gaussian $\delta$-correlated)

\begin{equation}\label{3.1}
\gamma(t)=2\beta\delta(t-t') \text{ where } \beta=c_0^2.
\end{equation}

\noindent Consequently the generalized Langevin equation (\ref{2.4})
reduces to

\begin{eqnarray}
\dot{x}&=&v\nonumber\\
\dot{v}&=&-\frac{dV(x)}{dx}-\beta(g'(x))^2v+g'(x)\{f(t)+\pi(t)\}\nonumber\\
&=&-\frac{dV(x)}{dx}-\Gamma(x)v+g'(x)\{f(t)+\pi(t)\}\label{3.2}
\end{eqnarray}

\noindent where

\begin{equation}\label{3.3}
\Gamma(x)=\beta(g'(x))^2.
\end{equation}

Using van Kampen's cumulant expansion method \cite{van1}, the
Fokker-Planck equation, valid for small correlation time,
corresponding to the above generalized Langevin equation is obtained
as (see Appendix A)

\begin{eqnarray}
\frac{\partial P(x,v,t)}{\partial t}& = & -v\frac{\partial
P}{\partial x} \nonumber \\
& & +[\Gamma(x)v+V'(x)-2g'(x)g''(x)J_e]\frac{\partial P}{\partial v}
\nonumber \\
& & + A\frac{\partial^2 P}{\partial v^2}+B\frac{\partial^2
P}{\partial v
\partial x}+\Gamma(x)P \label{n3.20}
\end{eqnarray}

\noindent where

\begin{equation}\label{n3.21}
A=(g'(x))^2I_e-\Gamma(x)(g'(x))^2J_e \ \text{ and } B=(g'(x))^2J_e ,
\end{equation}

\noindent and $I_e$ and $J_e$ are defined as

\begin{subequations}
\begin{eqnarray}
I_e &=& \int_0^\infty \langle \xi(t)\xi(t-\tau)\rangle d\tau, \label{n3.22a}\\
J_e &=& \int_0^\infty \tau \langle \xi(t)\xi(t-\tau)\rangle d\tau.
\label{n3.22b}
\end{eqnarray}
\end{subequations}

\noindent In equations (\ref{n3.22a}-\ref{n3.22b}), $\xi(t)$ is the
effective noise term [$\xi(t)=f(t)+\pi(t)$] as defined earlier. In
deriving (\ref{n3.20}) we have assumed that $f(t)$ and $\epsilon(t)$
are uncorrelated as they have different origin. \textit{Equation
(\ref{n3.20}) is the first key result of this paper}. It should be
noted that when the noise is purely internal and the
system-reservoir coupling is linear, equation (\ref{n3.20}) reduces
to the generalized Kramers equation \cite{han1} (valid for small
correlation time).

In Kramers' original treatment the dynamics of the Brownian particle
was governed by Markovian random processes. Since the work of
Kramers a number of authors
\cite{grote-hynes,hanggi-mojtabai,carmeli-nitzan,borkovec} have
extended Kramers' analysis for the non-Markovian case to derive the
expression for generalized escape rate. In order to allow ourselves
a comparison with Fokker-Planck equation of other forms
\cite{hanggi-mojtabai,carmeli-nitzan,adelman}, we note that the
diffusion coefficient in equation (\ref{n3.20}) is coordinate
dependent. It is customary to get rid of this coordinate dependence
by approximating the coefficients at the barrier top or potential
well where we need the steady state solution of equation
(\ref{n3.20}). One may also use mean field solution of equation
(\ref{n3.20}) obtained by neglecting the fluctuation terms and
putting appropriate stationary condition in the diffusion
coefficient. The drift term in equation (\ref{n3.20}) refers to the
presence of a dressed potential of the form

\begin{equation}
\label{3.23} R(x)=V(x)-(g'(x))^2J_e.
\end{equation}

\noindent The modification of the potential is essentially due to
nonlinear coupling of the system to the nonequilibrium modes. $J_e$
is a non-Markovian small contribution and therefore the second term
of the above equation may be neglected for small correlation time.
For the rest of the treatment we use $R(x)\simeq V(x)$. For a
harmonic oscillator with frequency $\omega_0$,
$V(x)=\omega_0^2x^2/2$; the linearized version of Fokker-Planck
equation is represented as

\begin{equation}\label{3.24}
\frac{\partial P}{\partial t}=-v\frac{\partial P}{\partial x}+\Gamma
P+[\Gamma v+\omega_0^2x]\frac{\partial P}{\partial
v}+A_0\frac{\partial^2 P}{\partial v^2}+B_0\frac{\partial^2
P}{\partial v\partial x}
\end{equation}

\noindent where

\begin{equation}\label{3.25}
A_0=(g'(0))^2I_e-\Gamma(0)(g'(0))^2J_e \text{ and } B_0=(g'(0))^2J_e
\end{equation}

\noindent are calculated at the bottom of the potential ($x=0$). From
equation (\ref{3.25}) we have

\begin{equation}\label{3.26}
A_0=(g'(0))^2I_e-\Gamma(0)B_0.
\end{equation}

\noindent The general steady state solution of equation (\ref{3.24})
becomes

\begin{equation}\label{3.27}
P_{st}(x,v)=\frac{1}{Z}\exp\left[ -\frac{v^2}{2D_0}-\frac{\omega_0^2
x^2}{2(D_0+B_0)} \right]
\end{equation}

\noindent where

\begin{equation}\label{3.28}
D_0=\frac{A_0}{\Gamma(0)}
\end{equation}

\noindent and $Z$ is the normalization constant. The solution
(\ref{3.27}) can be verified by direct substitution in the steady
state ($\partial P(x=0,v)/\partial t=0$) version of the Fokker-Planck
equation (\ref{n3.20}), namely

\begin{eqnarray}\label{3.29}
&& -v\frac{\partial P_{st}}{\partial x}+\Gamma P_{st}+[\Gamma
v+\omega_0^2x]\frac{\partial P_{st}}{\partial v} +
A_0\frac{\partial^2 P_{st}}{\partial v^2} \nonumber
\\
&& +B_0\frac{\partial^2 P_{st}}{\partial v\partial x}=0
\end{eqnarray}

\noindent The distribution (\ref{3.27}) is not an equilibrium
distribution. In absence of the external noise $\epsilon (t)$ it
reduces to the standard thermal Boltzmann distribution,
$\exp[-(v^2+V(x))/k_BT]$. Thus the steady state distribution for the
nonequilibrium open system plays the role of an equilibrium
distribution of the closed system which may however be recovered in
the absence of the external noise.

We now turn to the problem of decay of a metastable state. In
Kramers' approach \cite{kra}, the particle coordinate $x$
corresponds to the reaction coordinate, and its values at the minima
of the potential well $V(x)$ separated by a potential barrier,
denotes the reactant and product states.

Linearizing the motion around the barrier top at $x=x_b$, the steady
state ($\partial P(x=x_b,v)/\partial t=0$) Fokker-Planck equation
corresponding to equation (\ref{n3.20}) reads

\begin{eqnarray}\label{4.1}
&& -v\frac{\partial P_{st}}{\partial y}-\omega_b^2y\frac{\partial
P_{st}}{\partial v}+\Gamma(x_b)\frac{\partial}{\partial
v}(vP_{st})+A_b\frac{\partial^2 P_{st}}{\partial v^2} \nonumber
\\
&& +B_b\frac{\partial^2 P_{st}}{\partial v \partial y}=0 ,
\end{eqnarray}

\noindent where

\begin{equation}\label{4.2}
y=x-x_b, \text{ } V(y)=E_b-\frac{1}{2}\omega_b^2y^2, \text{ }
\omega_b^2>0
\end{equation}

\noindent and the suffix `$b$' indicates that all the coefficients
are to be calculated using the general definition of $A$ and $B$
(\ref{n3.21}) at the barrier top. It is interesting to note that for
linear coupling, we can extend our analysis for arbitrary
correlation time and in such a case, the barrier dynamics would have
been governed by the Fokker-Planck equation of Adelman's form
\cite{adelman}.

To derive a nonvanishing diffusion current across the barrier top
Kramers \cite{kra} considered $P_b(x,v)$ to be the equilibrium
Boltzmann distribution ($\exp[-(v^2+V(x))/k_BT]$) multiplied by a
propagator $F(x,v)$ and used it to solve the Fokker-Planck equation.
In our model the equilibrium distribution should be replaced by the
steady state distribution which depends on the local nature of the
potential at the barrier top and the effective temperature like
quantity for the nonequilibrium open system. However, in the absence
of the external noise $\epsilon(t)$ the steady state distribution
reduces to the equilibrium Boltzmann distribution. Following Kramers
\cite{kra}, we thus assume that the nonequilibrium steady state
probability $P_b(x,v)$ generating a nonvanishing diffusion current
across the barrier is given by

\begin{equation}\label{4.3}
P_b(x,v)=\exp\left[
-\left\{\frac{v^2}{2D_b}+\frac{V(x)}{D_b+B_b}\right\} \right]F(x,v)
,
\end{equation}

\noindent where

\begin{equation}\label{4.4}
D_b=\frac{A_b}{\Gamma(x_b)} ,
\end{equation}

\noindent with

\begin{eqnarray*}
V(x)&=&E_0+\frac{1}{2}\omega_0^2x^2, \text{ near the bottom}\\
&=&E_b+\frac{1}{2}\omega_b^2(x-x_b)^2, \text{ near the top} .
\end{eqnarray*}

\noindent The expression (\ref{4.3}) denoting the steady state
distribution is motivated by the local analysis near the bottom and
the top of the potential. For a stationary nonequilibrium system, on
the other hand, the relative population of the two regions, in
general, depends on the global properties of the potential leading
to an additional factor in the rate expression. Because of the
Kramers' type ansatz, which is valid for the local analysis, such a
consideration is outside the scope of the present treatment.

Following Kramers' original reasoning \cite{kra} we then derive the
barrier crossing rate $K$, for moderate to large friction regime
(see Appendix B for detailed calculation)

\begin{equation}\label{4.20}
K=\frac{\omega_0}{2\pi}\frac{D_b}{(D_0+B_0)^{1/2}}\left(\frac{\Lambda}{1+\Lambda
D_b }\right)^{1/2}\exp\left( \frac{-E_b}{D_b+B_b} \right)
\end{equation}

\noindent where $E_b$ is the barrier height of the potential and the
parameter $\Lambda$ is given by

\begin{equation*}
\Lambda=\frac{\lambda}{A_b+aB_b}
\end{equation*}

\noindent with

\begin{eqnarray*}
-\lambda & = & \Gamma(x_b)+a\left(1+\frac{B_b}{D_b}\right) ,\\
a & = &\frac{D_b}{2(D_b+B_b)}\left\{-\Gamma(x_b)-
\sqrt{\Gamma^2(x_b)+4\omega_b^2}\right\}.
\end{eqnarray*}

\noindent The strength of the external noise and the damping
function are buried in the parameters $D_0$, $B_0$, $D_b$, $B_b$ and
$\Lambda$. Equation (\ref{4.20}) is \textit{the second key result of
this paper}. Here we note that $(D_b+B_b)/k_B$ in the exponential
factor of (\ref{4.20}) defines a new effective temperature
characteristic of the steady state of the nonequilibrium open system
and an effective transmission factor is contained in the prefactor
controlling the barrier crossing dynamics. As expected both are the
functions of the external noise strength and the coupling of the
noise to the bath modes. In the absence of external stochastic
modulation, $\epsilon(t)=0$, equation (\ref{4.20}) reduces to
standard Kramers' result \cite{kra}, namely

\begin{equation*}
K_{\text{kramers}}=\frac{\omega_0}{2\pi\omega_b}\left[
\left\{\left(\frac{\beta}{2}\right)^2+\omega_b^2\right\}^{1/2}
-\frac{\beta}{2}\right]\exp\left(\frac{-E_b}{k_BT}\right)
\end{equation*}

\noindent
which can be verified with the explicit forms of the parameters
$D_0, B_0, D_b, B_b$ and $\Lambda$ with $J_e=0$, $I_e=\beta k_BT$,
$g(x)=x$ and $D$ or $\kappa_0$ equals to zero.


\begin{figure}[!t]
\includegraphics[width=1.0\linewidth,angle=0]{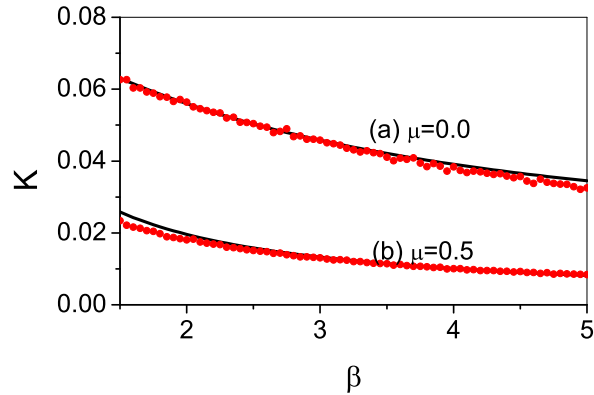}
\caption{\label{fig1} (color online) Variation of escape rate $K$
with dissipation constant $\beta$ with (a) $\mu=0$ and (b) $\mu=0.5$
using the parameter set $b_1=1.0$, $E_b=5.0$, $k_BT=0.1$,
$\kappa^2_0=5.0$, $D_e=0.1$ and $\tau_e=0.01$, where solid line and
circle represents analytical and numerical rates, respectively.}
\end{figure}


\begin{figure}[!b]
\includegraphics[width=1.0\linewidth,angle=0]{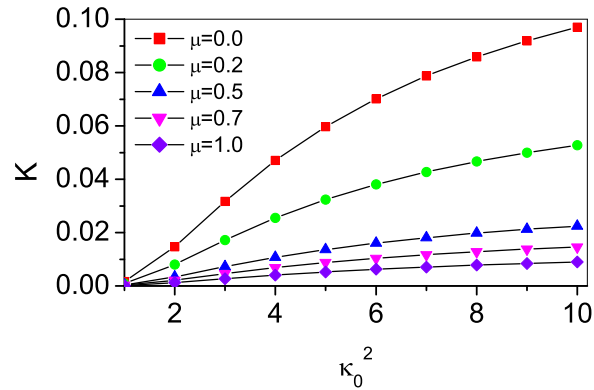}
\caption{\label{fig2}(color online) Variation of analytical rate
constant $K$ with $\kappa_0^2$ for different values of the strength
of nonlinearity $\mu$ with $b_1=1.0$, $E_b=5.0$, $k_BT=0.1$,
$\beta=1.0$, $D_e=1.0$ and $\tau_e=0.01$.}
\end{figure}

As the understanding of the theoretical aspects of nonequilibrium
statistical mechanics became better a vast body of literature have
emerged modifying Kramers' original approach which are well
documented in the review by H\"anggi, Talkner and Borkovec
\cite{han1}. In the following we briefly discuss one of the issues
from the post Kramers development of the reaction rate theory which
is closely connected to our work. Though Kramers' original approach
was restricted by the Markovian assumption, however, in certain
situations the memory effect becomes important and the generalized
Langevin equation with a memory kernel must be accounted for. To the
best of our knowledge following the seminal work in this direction
by Grote and Hynes \cite{grote-hynes}, H\"{a}nggi and Mojtabai
\cite{hanggi-mojtabai} and Carmeli and Nitzan \cite{carmeli-nitzan}
have extended Kramers' work for an arbitrary memory friction and
have found that the rate can often be larger than one would obtain
from Kramers' original approach. The analysis of H\"{a}nggi and
Mojtabai \cite{hanggi-mojtabai} was based on the non-Markovian
Fokker-Planck equation of Adelman \cite{adelman} for a parabolic
potential and they have used essentially the same approach of
Kramers. The generalized Fokker-Planck approach have also been
adopted by Carmeli and Nitzan \cite{carmeli-nitzan} to derive the
expression for the steady state escape rate in the non-Markovian
regime. All the above mentioned theoretical approaches have been
first supplemented with a full stochastic simulation by
Straub,Borkovec and Berne \cite{borkovec} where the authors
explicitly studied the dynamics within the framework of
non-Markovian generalized Langevin equation. Incorporation of memory
effect in the above mentioned work gets reflected in the rate
expression

\begin{equation*}
K_{\text{memory}}=\frac{\omega_0}{2\pi\omega_b}\left[
\left\{\left(\frac{\overline{\gamma}}{2}\right)^2+
\overline{\omega}_b^2\right\}^{1/2}
-\frac{\overline{\gamma}}{2}\right]\exp\left(\frac{-E_b}{k_BT}\right)
\end{equation*}

\noindent where, $\overline{\gamma}$ and $\overline{\omega}_b$ are
long time limit of the memory kernel $\gamma (t)$ and the
renormalized frequency at the top of the potential $\omega_b(t)$,
respectively. In this paper we have extended the above mentioned
approaches \cite{hanggi-mojtabai,carmeli-nitzan,adelman} for state
dependent diffusion in the non-Markovian regime to obtain a
generalized steady state escape rate when the bath is modulated by
an external stochastic force. Though our treatment is valid for
small correlation time, it incorporates most of the characteristics
of the non-Markovian state dependent diffusion process.



\begin{figure}[!t]
\includegraphics[width=1.0\linewidth,angle=0]{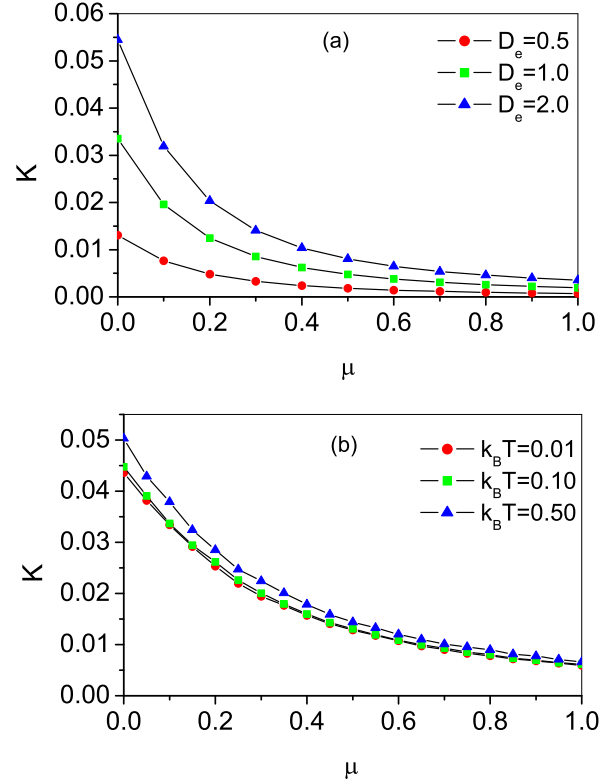}
\caption{\label{fig3} (color online) Variation of rate constant $K$
with strength of nonlinearity $\mu$ (a) for different values of
external noise strength $D_e$ (calculated analytically using
$k_BT=0.1$) and (b) for different values of temperature $k_BT$
(calculated numerically using $D_e=1.0$). The values of the other
parameters used are $b_1=1.0$, $E_b=5.0$, $\beta=3.0$,
$\kappa^2_0=5.0$ and $\tau_e=0.01$.}
\end{figure}

\section{Specific example: Heat bath driven by external colored noise}

As a specific example we consider that the heat bath is modulated
externally by a colored noise $\epsilon(t)$ with noise correlation

\begin{equation}\label{5.1}
\langle \epsilon(t)\epsilon(t')
\rangle_e=\frac{D_e}{\tau_e}\exp\left(-\frac{|t-t'|}{\tau_e}\right)
\end{equation}

\noindent where $D_e$ and $\tau_e$ are the strength and correlation
time of the external noise respectively. In addition to that we
consider the internal noise $f(t)$ to be white (\textit{i.e.}
$\tau_c \rightarrow 0$). The effective Gaussian-Ornstein-Uhlenbeck
noise $\xi(t)=f(t)+\pi(t)$ will have an intensity $D_R$ and a
correlation time $\tau_R$ given by \cite{lin}

\begin{eqnarray*}
D_R &=& \int_0^\infty \langle \xi(t)\xi(0) \rangle dt ,\\ \tau_R
&=&\frac{1}{D_R}\int_0^\infty t \langle\xi(t)\xi(0)\rangle dt .
\end{eqnarray*}

\noindent Following the above definitions and using equation (\ref{2.9})
we have (see equation (\ref{2.17})),

\begin{subequations}
\begin{eqnarray}
D_R&=&c_0^2(k_BT+D_e\kappa_0^2)=\beta(k_BT+D_e\kappa_0^2)\label{5.2a} ,\\
\tau_R&=&\frac{D_e c_0^2 \kappa_0^2}{D_R}\tau_e=\frac{\beta D_e
\kappa_0^2}{D_R}\tau_e\label{5.2b} .
\end{eqnarray}
\end{subequations}

\noindent Consequently the functions $A(x)$ and $B(x)$ in
equation (\ref{n3.20}) becomes

\begin{subequations}
\begin{eqnarray}
A(x)&=&(g'(x))^2D_R-\Gamma(x)(g'(x))^2\tau_R D_R\label{5.3a} ,\\
B(x)&=&(g'(x))^2\tau_R D_R\label{5.3b} .
\end{eqnarray}
\end{subequations}

\noindent From these equations we may evaluate the various
parameters to obtain the generalized escape rate from
equation (\ref{4.20}).



\begin{figure}[t!]
\includegraphics[width=1.0\linewidth,angle=0]{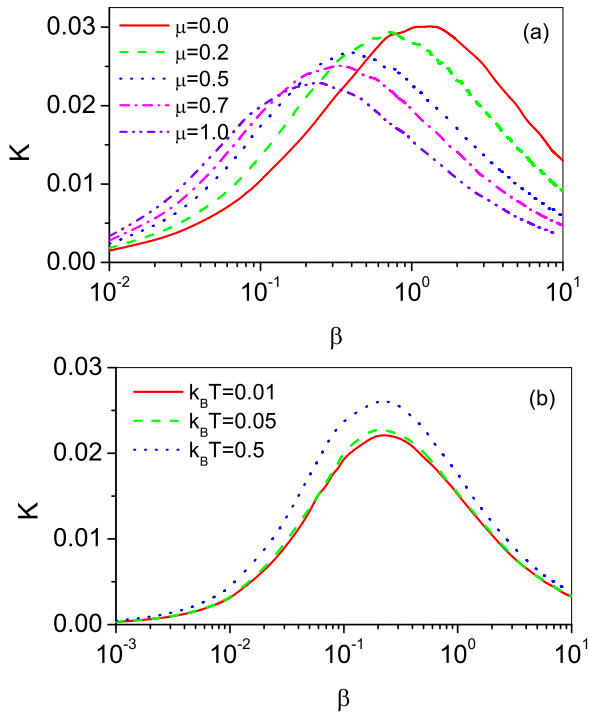}
\caption{\label{fig4} (color online) Turnover of rate constant $K$
(numerical) with dissipation constant $\beta$ for (a) different
values of $\mu$ (for $k_BT=0.1$) and (b)different values of $k_BT$
(for $\mu=0.5$). The values of other parameters used are $b_1=1.0$,
$E_b=5.0$, $D_e=1.0$, $\kappa^2_0=5.0$, $\tau_e=1.0$. }
\end{figure}

\section{Results and discussions}

To study the dynamics we consider a model cubic potential of the
form $V(x)=b_1x^2-b_2x^3$ where $b_1$ and $b_2$ are the two
constant parameters with $b_1, b_2>0$, so that the activation energy
becomes $E_b=4b_1^3/27b_2^2$. The nonlinear coupling function is
taken to be $g(x)=x+(1/2)\mu x^2$, $\mu$ being a constant implying
the strength of nonlinearity of the coupling function. We then
numerically solve the Langevin equation (\ref{3.2}) using second
order stochastic Heun's algorithm. To ensure the stability of our
simulation we have used a small time step $\Delta t=0.001$, with
$\Delta t/\tau_R \ll 1$. The numerical rate has been defined as the
inverse of the mean first passage time \cite{mah}. The mean first
passage time has been calculated by averaging over $10,000$
trajectories. The value of other parameters used are given in the
respective figure.

One of the result of Kramers' theory is that $K$ varies inversely in
the intermediate to strong damping regime. In Fig.1 we have plotted
the rate constant $K$ against the damping constant $\beta=c_0^2$ in
the moderate to large damping region where our theory is valid and
we compare the theoretical result (\ref{4.20}) with the numerical
simulation data for two different values of the nonlinear coupling
parameter $\mu$. It is observed that the agreement between the
theoretical prediction and numerical simulation is quite
satisfactory. In Fig.2 we plot the variation of rate constant $K$,
obtained from theoretical result (\ref{4.20}), with external
coupling constant $\kappa_0^2$ for various nonlinearity parameters.
We observe that for a given $\mu$, the rate increases nearly
linearly and for a particular value of $\kappa_0$, increase in $\mu$
causes decrease in rate, which is also observed from the Fig.3(a)
(where we have plotted $K$ vs. $\mu$ for different values of $D_e$
from analytical result) and Fig.3(b) (where the same has been
observed numerically for different temperatures).

In his dynamical theory of chemical reaction, Kramers identified two
distinct regimes of stationary nonequilibrium states in terms of
dissipation constant $\beta$. The essential result of Kramers'
theory is that the rate varies linearly in the weak damping regime
and inversely in the intermediate to strong damping regime. In
between the two regimes the rate constant as a function of damping
constant exhibits a bell-shaped curve known as Kramers' turnover
\cite{han1}. In the traditional system reservoir model the
dissipation and fluctuations are connected through the
fluctuation-dissipation relation. A typical signature of this
relation can been seen through this turnover phenomenon. Whereas for
a thermodynamically open system where the heat bath is modulated by
external noise, both the dissipation $\beta$ and response function
$\varphi$ depend on the properties of the reservoir. Due to this
connection between the dissipation and external noise source
equation (\ref{2.9}), plays the typical role of thermodynamic consistency
relation, an analog of the fluctuation-dissipation relation for a
thermodynamically closed system, for which one can expect turnover
feature for this open system. In Fig.4(a) and Fig.4(b) we have
plotted the rate constant $K$ obtained from Langevin simulation for
a wide range of damping constant for different values of
nonlinearity parameter $\mu$ and temperature, respectively. The
figures apart form demonstrating the turnover of the rate constant
with the variation of the damping constant, shows a shifting of
maxima towards left \textit{i.e.}, weak damping regime with the
increase of $\mu$ and also is consistent with Fig.3.

While observing the variation of the rate constant $K$ as a function
of correlation time ($\tau_e$) of the external noise for different
$D_e$ (Fig.5(a)), different temperature (Fig.5(b)) and for different
nonlinearity parameter $\mu$ (Fig.5(c)) we find an interesting
result. In all cases $K$ passes through a maxima, then decreases and
ultimately becomes independent of $\tau_e$ for large values of
$\tau_e$. In short the rate of barrier crossing exhibits a resonance
behavior with the correlation time of external noise which is
responsible for fluctuation of barrier height. The above resonance
phenomenon is known as resonant activation (RA) \cite{ra}. So far RA
has been observed due to the barrier fluctuation as a result of
direct driving of correlated noise to the system. In this model the
barrier fluctuation occurs (\ref{3.2}) due to the driving of
nonlinearly coupled heat bath with the system by correlated noise.
So the RA occurs as a result of correlated noise driven bath, which
is interesting and new feature, instead of direct system driving.
The immediate experimentally observable situation could be if we
consider a simple unimolecular isomerization reaction in a
photochemically active solvent under the influence of fluctuating
light intensity (see section I), the reaction rate can be enhanced
by tuning the correlation time of the fluctuating light field. It is
also interesting to note that our nonlinear coupling model which
yields a state dependent diffusion may have an important consequence
in the generation of current for a Brownian particle moving in a
periodic potential without any external bias. Because of its
extraordinary success in explaining experimental observations on
biomolecular motors active in muscle contractions, the state
dependent diffusion has attracted wide attentions in recent years
\cite{ast}.



\begin{figure}[t!]
\includegraphics[width=1.0\linewidth,angle=0]{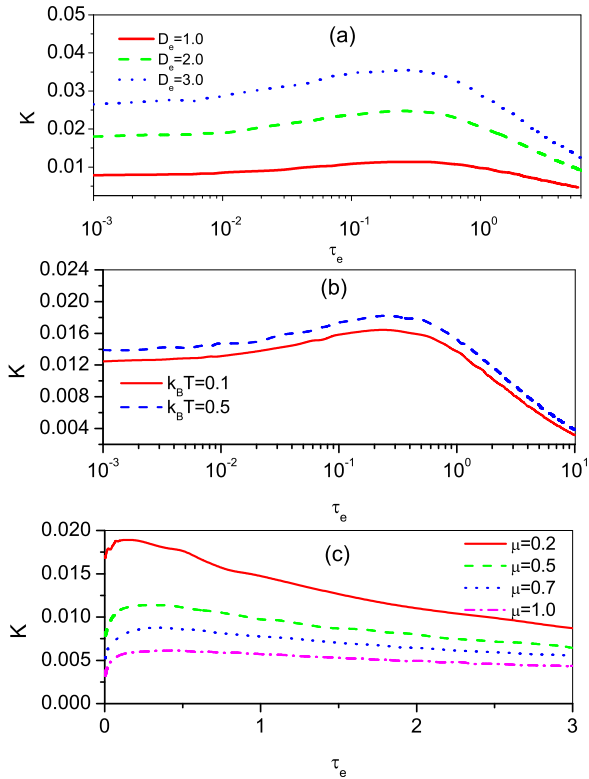}
\caption{\label{fig5} (color online) Numerical variation of rate
constant $K$ with correlation of external noise $\tau_e$ for (a)
different values of strength of external noise $D_e$ (for $k_BT=0.1$
and $\mu=0.5$), (b) different values of $k_BT$ (for $D_e=1.0$ and
$\mu=0.5$) and (c) different values of $\mu$ (for $k_BT=1.0$ and
$D_e=1.0$). The values of other parameters used are $b_1=1.0$,
$E_b=5.0$, $\beta=5.0$ (for (b) $\beta=3.0$) and $\kappa^2_0=5.0$.
Note the logarithmic abscissa in (a) and (b).}
\end{figure}

\section{Conclusion}

Based on a system reservoir microscopic model where the system is
nonlinearly coupled to a heat bath which is modulated by an
external, stationary Gaussian noise with an arbitrary decaying
correlation function, we have generalized the Kramers' theory and
have numerically analyzed the model to calculate the steady state
rate of escape from a metastable well. We have shown that since the
reservoir is driven by the external noise and the dissipative
properties of the system depend on the reservoir, a simple
connection between the dissipation and the response function of the
medium due to external noise can be established. We then followed
the dynamics of the open system in the spatially-diffusion limited
regime and derived the Fokker-Planck equation (corresponding to the
Langevin equation with space dependent dissipation and
multiplicative noise) with space dependent diffusion coefficient
containing the effective temperature like quantity. We then derived
the generalized Kramers' escape rate for moderate to strong damping
regime. From the point of view of the realistic situation we
considered the special case where the internal noise is white and
the external noise is colored and have calculated the escape rate
for a model cubic potential. We also numerically simulate the
Langevin equation and observe that the theoretical prediction agrees
reasonably well with the numerical result. Dependence of the rate
$K$ on various other parameters have been studied and it is observed
that the enhancement of rate is possible by tuning the correlation
time of external noise. The creation of a typical nonequilibrium
open situation by modulating a bath with the help of an external
noise is not an uncommon phenomena in applications and industrial
processing. The external agency generating noise does work on the
bath by stirring, pumping, agitating etc to which the system
dissipates internally. In the present treatment we are concerned
with a nonequilibrium steady state that signifies a constant
throughput of energy. We believe that these considerations are
likely to be important in other related issues in nonequilibrium
open systems such as thermal ratchet and related problems.

So far, in this paper, we have considered a linear coupling in the
interaction between the heat bath and the external driving noise,
$H_{int}$. It will be interesting to see how the dynamics changes
when $H_{int}$ is nonlinear. Studies have been made recently on
anomalous diffusion in presence of correlated external noise
\cite{den}. Our present methodology may be extended to investigate
the transport process when the bath is modulated by two correlated
external noise. In our future communication we would like to address
such issues.

\begin{acknowledgments}
We are thankful to Professor D.S. Ray for stimulating discussions and
critical comments. JRC is also thankful to Dr. B. Deb for constant
encouragement and to Professor J.K. Bhattacharjee for useful
discussions. JRC is indebted to Indian Academy of Science for
providing a summer fellowship. SKB acknowledges support from
Virginia Tech through ASPIRES award program.
\end{acknowledgments}


\appendix

\section{Derivation of Fokker-Planck equation}

Equation (\ref{3.2}) can be written in the following form

\begin{subequations}
\begin{eqnarray}
\dot{u_1}&=&G_1(u_1,u_2,t;f(t),\pi(t))\label{3.4a}\\
\dot{u_2}&=&G_2(u_1,u_2,t;f(t),\pi(t))\label{3.4b}
\end{eqnarray}
\end{subequations}

\noindent where we have used the following abbreviations

\begin{subequations}
\begin{eqnarray}
u_1&=&x\label{3.5a}\\
u_2&=&v\label{3.5b}
\end{eqnarray}
\end{subequations}

\noindent and

\begin{subequations}
\begin{eqnarray}
G_1&=&v\label{3.6a}\\
G_2&=&-\frac{dV(x)}{dx}-\Gamma(x)v+g'(x)\{f(t)+\pi(t)\}\label{3.6b}
.
\end{eqnarray}
\end{subequations}

\noindent The vector $u$ with components $u_1$ and $u_2$ thus
represents a point in a two-dimensional ``phase space'' and the
equation (\ref{3.4a}-\ref{3.4b}) determines the velocity at each
point in this phase space. The conservation of the points now
asserts the following linear equation of motion for density
$\rho(u,t)$ in phase space \cite{van1}

\begin{equation}\nonumber
\frac{\partial}{\partial
t}\rho(u,t)=-\sum_{n=1}^2\frac{\partial}{\partial
u_n}G_n(u,t,f(t),\pi(t))\rho(u,t)
\end{equation}

\noindent or more compactly

\begin{equation}\label{3.7}
\frac{\partial \rho}{\partial t}=-\nabla\cdot G \rho.
\end{equation}

Our next task is to find out a differential equation whose average
solution is given by $\langle \rho \rangle$ \cite{van1} where the
stochastic averages has to be performed over two noise processes
$f(t)$ and $\epsilon(t)$. $\nabla\cdot G$ can be partitioned into
two parts: a constant part $\nabla\cdot G_0$ and a fluctuating part
$\nabla\cdot G_1(t)$, containing these noises. Thus we write

\begin{equation}\label{3.8}
\nabla\cdot G(u,t,f(t),\pi(t))=\nabla\cdot G_0(u)+\alpha \nabla\cdot
G_1(u,t,f(t),\pi(t))
\end{equation}

\noindent where $\alpha$ is a parameter (we put it as an external
parameter to keep track of the order of the perturbation expansion
in $\alpha\tau_e$, $\tau_e$ is the correlation time of the
$\epsilon(t)$, we put $\alpha=1$ at the end of the calculation) and
also note that $\langle\langle G_1(t) \rangle\rangle=0$.
Equation (\ref{3.7}) therefore takes the following form

\begin{equation}\label{3.9}
\dot{\rho}(u,t)=(A_0+\alpha A_1)\rho(u,t)
\end{equation}

\noindent where $A_0=-\nabla \cdot G_0$ and $A_1=-\nabla \cdot G_1$.
The symbol $\nabla$ is used for the operator that differentiate
everything that comes after it with respect to $u$. Making use of
one of the main results for the theory of linear equation of the
form (\ref{3.9}) with multiplicative noise \cite{van1}, we derive an
average equation for $\rho$ [$\langle \rho \rangle=P(u,t)$, the
probability density of $u(t)$]

\begin{eqnarray}\label{3.10}
\frac{\partial P}{\partial t}& = & \left\{
A_0+\alpha^2\int_0^{\infty}d\tau \langle A_1(t)\exp(\tau
A_0)A_1(t-\tau) \rangle \right. \nonumber \\
& & \left. \times \exp(-\tau A_0) \right\}P.
\end{eqnarray}

\noindent The above result is based on second order cumulant
expansion and is valid for the rapid fluctuations with small
strength where the correlation time $\tau_e$ is short but finite
\cite{van1}, \textit{i.e.},

\begin{equation}\nonumber
\langle A_1(t)A_1(t')\rangle=0 \text{ for } |t-t'|>\tau_e .
\end{equation}

\noindent Equation (\ref{3.10}) is exact in the limit $\tau_e$
trends to zero. Using the expansions for $A_0$ and $A_1$ we obtain

\begin{eqnarray}\label{3.11}
\frac{\partial P}{\partial t} & = & \left\{ -\nabla\cdot
G_0+\alpha^2\int_0^{\infty}d\tau \langle \nabla\cdot
G_1(t)\exp(-\tau\nabla\cdot G_0) \right. \nonumber \\
&& \left. \times \nabla\cdot G_1(t-\tau) \rangle \exp(\tau
\nabla\cdot G_0) \right\}P .
\end{eqnarray}

\noindent The operator $\exp(-\tau \nabla\cdot G_0)$ in the above
equation provides the solution of the equation

\begin{equation}\label{3.12}
\frac{\partial \mathscr{G}(u,t)}{\partial t}=-\nabla \cdot G_0
\mathscr{G}(u,t)
\end{equation}

\noindent ($\mathscr{G}$ signifies the unperturbed part of $\rho$)
which can be found explicitly in terms of characteristic curves. The
equation

\begin{equation}\label{3.13}
\dot{u}=G_0(u)
\end{equation}

\noindent for fixed $t$ determines a mapping from $u(\tau=0)$ to
$u(\tau)$, \textit{i.e.}, $u\rightarrow u^{\tau}$ with the inverse
$(u^\tau)^{-\tau}=u$. The solution of equation (\ref{3.12}) is given
by

\begin{equation}\label{3.14}
\mathscr{G}(u,t)=\mathscr{G}(u^{-t},0) \left |\frac{d(u^{-t})}{d(u)}
\right |=\exp(-t\nabla\cdot G_0)\mathscr{G}(u,0) ,
\end{equation}

\noindent $|d(u^{-t})/d(u)|$ being a Jacobian determinant. The
effect of $\exp(-t\nabla\cdot G_0)$ on $\mathscr{G}(u)$ is given by

\begin{equation}\label{3.15}
\exp(-t\nabla\cdot G_0) \mathscr{G}(u,0)=\mathscr{G}(u^{-t},0) \left
|\frac{d(u^{-t})}{d(u)} \right | .
\end{equation}

\noindent The above simplification when we put in equation
(\ref{3.11}) yields

\begin{eqnarray}\label{3.16}
\frac{\partial P}{\partial t} & = & \nabla \cdot \left\{
-G_0+\alpha^2 \int_0^\infty d\tau \left
|\frac{d(u^{-\tau})}{d(u)}\right | \right. \nonumber \\
&& \times \langle G_1(u,t) \nabla_{-\tau}\cdot G_1(u^{-\tau},t-\tau)
\rangle \nonumber \\
&& \left. \times \left | \frac{d(u)}{d(u^{-\tau})}\right |\right\} P
.
\end{eqnarray}

\noindent where $\nabla_{-\tau}$ denotes differentiation with
respect to $u_{-\tau}$. We put $\alpha=1$ for the rest of the
treatment. We now identify

\begin{eqnarray}
u_1 &=& x \nonumber \\
u_2 &=& v \nonumber \\
G_{01} &=& v, \text{ } G_{11}=0 \label{3.17} \\
G_{02} &=& -\Gamma(x)v-V'(x) \nonumber \\
G_{12} &=& g'(x)\{f(t)+\pi(t)\}\nonumber.
\end{eqnarray}

\noindent In this notation equation (\ref{3.16}) now reduces to

\begin{eqnarray}\label{3.18}
\frac{\partial P}{\partial t}& = &-\frac{\partial}{\partial x}(v
P)+\frac{\partial}{\partial v}\{\Gamma(x)v+V'(x)\}P\nonumber\\
& & + \frac{\partial}{\partial v}\int_0^\infty d\tau
\langle[g'(x)\{f(t)+\pi(t)\}] [\frac{\partial}{\partial v^{-\tau}}
\nonumber \\
& & \times \{g'(x^{-\tau})(f(t-\tau)+\pi(t-\tau))\}]\rangle P
\end{eqnarray}

\noindent where we have used the fact that the Jacobian obeys the
equation \cite{van1}

\begin{equation}\nonumber
\frac{d}{dt}\log
\left|\frac{d(x^t,v^t)}{d(x,v)}\right|=\frac{\partial v}{\partial
x}+\frac{\partial}{\partial v}\left\{-\Gamma v+V'(x)\right\}=-\Gamma
,
\end{equation}

\noindent so that the Jacobian equals to $e^{-\Gamma t}$.

As a next approximation we consider the``unperturbed'' part of
equation (\ref{3.4a}-\ref{3.4b}) and take the variation of $v$ during
$\tau_e$ into account to first order in $\tau_e$. Thus we have

\begin{equation}\label{3.19}
x^{-\tau}=x-\tau v, \text{ } v^{-\tau}=v+\Gamma\tau v+\tau V'(x).
\end{equation}

\noindent Neglecting terms $\mathscr{O}(\tau^2)$, equation (\ref{3.19})
yields

\begin{equation}\nonumber
\frac{\partial}{\partial v^{-\tau}}=(1-\Gamma
\tau)\frac{\partial}{\partial v}+\tau \frac{\partial}{\partial x}.
\end{equation}

Taking this into consideration equation (\ref{3.18}) can be simplified
after some algebra to the following form

\begin{eqnarray}
\frac{\partial P(x,v,t)}{\partial t}& = & -v\frac{\partial
P}{\partial x} \nonumber \\
& & +[\Gamma(x)v+V'(x)-2g'(x)g''(x)J_e]\frac{\partial P}{\partial v}
\nonumber \\
& & + A\frac{\partial^2 P}{\partial v^2}+B\frac{\partial^2
P}{\partial v
\partial x}+\Gamma(x)P \label{3.20}
\end{eqnarray}

\noindent where

\begin{equation}\label{3.21}
A=(g'(x))^2I_e-\Gamma(x)(g'(x))^2J_e \ \text{ and } B=(g'(x))^2J_e ,
\end{equation}

\noindent and $I_e$ and $J_e$ are defined as

\begin{subequations}
\begin{eqnarray}
I_e &=& \int_0^\infty \langle \xi(t)\xi(t-\tau)\rangle d\tau, \label{3.22a}\\
J_e &=& \int_0^\infty \tau \langle \xi(t)\xi(t-\tau)\rangle d\tau.
\label{3.22b}
\end{eqnarray}
\end{subequations}

\section{Derivation of escape rate}

Inserting (\ref{4.3}) in (\ref{4.1}), we obtain the equation for
$F(x,v)$ in the steady state in the neighborhood of $x_b$

\begin{eqnarray}
& - & (1+B_b/D_b)v\frac{\partial F}{\partial
x}-\left[\frac{D_b}{D_b+B_b}\omega_b^2(x-x_b) \right. \nonumber
\\
& + & \left. \Gamma(x_b)v\right]\frac{\partial F}{\partial v} +
A_b\frac{\partial^2F}{\partial v^2}+B_b\frac{\partial^2 F}{\partial
v \partial x}=0 . \label{4.5}
\end{eqnarray}

\noindent At this point we set

\begin{equation}\label{4.6}
u=v+a(x-x_b) ,
\end{equation}

\noindent where $a$ is a constant to be determined. With the help of
transformation (\ref{4.6}), equation (\ref{4.5}) reduces to the following
form

\begin{eqnarray}
& &
\{A_b+aB_b\}\frac{d^2F}{du^2}-\left[\frac{D_b}{D_b+B_b}\omega_b^2(x-x_b)
\right. \nonumber \\
& & \left. +
\left\{\Gamma(x_b)+a\left(1+\frac{B_b}{D_b}\right)\right\}v\right]\frac{dF}{du}=0.
\label{4.7}
\end{eqnarray}

\noindent Now let

\begin{equation}\label{4.8}
\frac{D_b}{D_b+B_b}\omega_b^2 (x-x_b)+
\left\{\Gamma(x_b)+a\left(1+\frac{B_b}{D_b}\right)\right\}v=-\lambda
u
\end{equation}

\noindent where $\lambda$ is another constant to be determined
later. By virtue of the relation (\ref{4.8}), equation (\ref{4.7}) becomes

\begin{equation}\label{4.9}
\frac{d^2F}{du^2}+\Lambda u\frac{dF}{du}=0
\end{equation}

\noindent where

\begin{equation}\label{4.10}
\Lambda=\frac{\lambda}{A_b+aB_b} .
\end{equation}

\noindent The two constants $\lambda$ and $a$ must satisfy the
simultaneous relations

\begin{eqnarray*}
-\lambda a&=&\frac{D_b}{D_b+B_b}\omega_b^2 ,\\
-\lambda &=&\Gamma(x_b)+a\left(1+\frac{B_b}{D_b}\right) .
\end{eqnarray*}

\noindent This implies that the constant $a$ must satisfy the
quadratic equation

\begin{equation}\nonumber
\frac{D_b+B_b}{D_b}a^2+\Gamma(x_b)a-\frac{D_b}{D_b+B_b}\omega_b^2=0
\end{equation}

\noindent which allows the solutions for $a$ as

\begin{equation}\label{4.11}
a_{\pm}=\frac{D_b}{2(D_b+B_b)}\left\{-\Gamma(x_b)\pm
\sqrt{\Gamma^2(x_b)+4\omega_b^2}\right\} .
\end{equation}

\noindent The general solution of equation (\ref{4.9}) is

\begin{equation}\label{4.12}
F(u)=F_2\int_0^u \exp\left(-\frac{\Lambda z^2}{2}\right) dz+F_1 ,
\end{equation}

\noindent where $F_1$ and $F_2$ are constant of integration. We look
for a solution which vanishes for large $x$. This condition is
satisfied if the integration in (\ref{4.12}) remain finite for
$|u|\rightarrow + \infty$. This implies that $\Lambda > 0$ so that
only $a_{-}$ becomes relevant. Then the requirement
$P_b(x,v)\rightarrow 0$ for $x\rightarrow +\infty$ yields

\begin{equation}\label{4.13}
F_1=F_2\sqrt{\pi/2\Lambda}.
\end{equation}

\noindent Thus we have

\begin{equation}\nonumber
F(u)=F_2\left[\sqrt{\frac{\pi}{2\Lambda}}+\int_0^u
\exp\left(-\frac{\Lambda z^2}{2}\right) dz\right]
\end{equation}

\noindent and correspondingly

\begin{eqnarray}
P_b(x,v)& = & F_2\left[\sqrt{\frac{\pi}{2\Lambda}}+\int_0^u
\exp\left(-\frac{\Lambda z^2}{2}\right) dz\right] \nonumber
\\
& & \times \exp\left[
-\left\{\frac{v^2}{2D_b}+\frac{V(x)}{D_b+B_b}\right\} \right].
\label{4.14}
\end{eqnarray}

The current across the barrier associated with the steady state
distribution is given by

\begin{equation}\nonumber
j=\int_{-\infty}^{+\infty} v P_b(x=x_b, v) dv
\end{equation}

\noindent which may be evaluated using (\ref{4.14}) and the
linearized version of $V(x)$, namely
$V(x)=E_b-(1/2)\omega_b^2(x-x_b)^2$ as

\begin{equation}\label{4.15}
j=F_2\left(\frac{2\pi}{\Lambda+D_b^{-1}}\right)^{1/2} D_b \exp\left(
\frac{-E_b}{D_b+B_b} \right) .
\end{equation}

To determine the remaining constant $F_2$ we proceed as follows. We
first note that as $x\rightarrow -\infty$ the pre-exponential factor
in equation (\ref{4.14}) reduces to the following form

\begin{equation}\label{4.16}
F_2[...]=F_2\left(\frac{2\pi}{\Lambda}\right)^{1/2} .
\end{equation}

\noindent We then obtain the reduced distribution function in $x$ as

\begin{equation}\label{4.17}
\widetilde{P}_b(x\rightarrow -\infty)=2\pi
F_2\left(\frac{D_b}{\Lambda}\right)^{1/2}\exp\left(
\frac{-V(x)}{D_b+B_b} \right) ,
\end{equation}

\noindent where we have used the definition for the reduced
distribution as

\begin{equation}\nonumber
\widetilde{P}(x)=\int_{-\infty}^{+\infty} P(x,v) dv .
\end{equation}

\noindent Similarly we derive the reduced distribution in the left
well around $x\approx 0$ using equation (\ref{3.27}) where the linearized
potential is $V(x)=\omega_0^2 x^2/2$,

\begin{equation}\label{4.18}
\widetilde{P}_{st}(x)=\frac{1}{Z}\sqrt{2\pi D_0}\exp\left(
\frac{-\omega_0^2x^2}{2(D_0+B_0)} \right)
\end{equation}

\noindent with the normalization constant $1/Z$ given by

\begin{equation}\nonumber
\frac{1}{Z}=\frac{\omega_0}{2\pi \sqrt{D_0(D_0+B_0)}} .
\end{equation}

The comparison of the distribution (\ref{4.17}) and (\ref{4.18})
near $x\approx 0$, \textit{i.e.},

\begin{equation}\nonumber
\widetilde{P}_{st}(x_0)=\widetilde{P}_b(x_0)
\end{equation}

\noindent gives

\begin{equation}\label{4.19}
F_2=\left(\frac{\Lambda}{D_b}\right)^{1/2}\frac{\omega_0}{2\pi
\sqrt{2\pi (D_0+B_0)}}.
\end{equation}

\noindent Hence from equation (\ref{4.15}), the normalized current
or the barrier crossing rate $K$, for moderate to large friction, is
given by

\begin{equation}
K=\frac{\omega_0}{2\pi}\frac{D_b}{(D_0+B_0)^{1/2}}\left(\frac{\Lambda}{1+\Lambda
D_b }\right)^{1/2}\exp\left( \frac{-E_b}{D_b+B_b} \right)
\end{equation}

\noindent where $E_b$ is the potential barrier height.



\begin{thebibliography}{99}

\bibitem{expt1} E.W.-G. Diau, J.L. Herek, Z.H. Kim, and A.H.
Zewail, Science \textbf{279}, 847 (1998).

\bibitem{expt2} L.I. McCann, M. Dykman, and B. Golding, Nature
\textbf{402}, 785 (1999).

\bibitem{kra} H. A. Kramers, Physica (Amsterdam) \textbf{7}, 284
(1940).

\bibitem{han1} P. H\"{a}nggi, P. Talkner, and M. Borkovec, Rev.
Mod. Phys. \textbf{62}, 251 (1990); V. I. Mel'nikov, Phys. Reps.
\textbf{209}, 1 (1991).

\bibitem{lan} R. Landauer and J. A. Swanson, Phys. Rev. \textbf{121},
1668 (1961); J. S. Langer, Ann. Phys. (N.Y.) \textbf{54}, 258
(1969).

\bibitem{tak} P. Talkner and D. Ryter, Phys. Lett. A \textbf{88}, 162
(1982).

\bibitem{van} N. G. van Kampen, Prog. Theo. Phys. Supl. \textbf{64}, 389
(1978).

\bibitem{han2} P. H\"{a}nggi, Phys. Lett. A \textbf{78}, 304
(1980).

\bibitem{ski} J. L. Skinner and P. G. Wolynes,
J. Chem. Phys. \textbf{69}, 2143 (1978); \textit{ibid} \textbf{72}, 4913 (1980).

\bibitem{jrc1} J. Ray Chaudhuri, G. Gangopadhyay and D. S. Ray,
J. Chem. Phys. \textbf{109}, 5565 (1998); M. M. Millonas and C. Ray,
Phys. Rev. Lett. \textbf{75}, 1110 (1995).

\bibitem{wol} P. G. Wolynes, Phys. Rev. Lett. \textbf{47}, 968 (1987);
W. H. Miller, J. Chem. Phys. \textbf{62}, 1899 (1975); A. O.
Caldeira and A. J. Leggett, Phys. Rev. Lett. \textbf{46}, 211
(1981); H. Grabert, P. Schramm and G. L. Ingold, Phys. Rep. \textbf{
168}, 115 (1988).

\bibitem{jrc2} J. Ray Chaudhuri, B. C. Bag and D. S. Ray, J. Chem. Phys.
\textbf{111}, 10852 (1999).

\bibitem{dsr} D. Banerjee, B. C. Bag, S. K.
Banik and D. S. Ray, J. Chem. Phys. \textbf{120}, 8960 (2004); D.
Barik, S. K. Banik and D. S. Ray, J. Chem. Phys. \textbf{119}, 680
(2003); D. Barik and D. S. Ray, J. Stat. Phys. \textbf{120}, 339
(2005).

\bibitem{hor} W. Horsthemke and R. Lefever, \textit{Noise-Induced
Transitions} (Springer-Verlag, Berlin, 1984).

\bibitem{san1} J. M. Sancho, M. San Miguel, S. L. Katz and J. D.
Gunton, Phys. Rev. A \textbf{26}, 1589 (1982) and references
therein.

\bibitem{lin} K. Lindenberg and B. J. West,
\textit{The Nonequilibrium Statistical Mechanics of Open and Closed
Systems} (VCH Publisher, Inc., New York, 1990).

\bibitem{jrc3} J. Ray Chaudhuri, D. Barik and S. K. Banik, Phys. Rev. E.
\textbf{73}, 051101 (2006); J. Ray Chaudhuri, S. K. Banik, B. C. Bag
and D. S. Ray, Phys. Rev. E \textbf{63}, 061111 (2001).

\bibitem{aus} R. D. Astumian, Science \textbf{276}, 917 (1997); P. Reimann,
Phys. Rep. \textbf{361}, 57 (2002).

\bibitem{jrc4} S. K. Banik, J. Ray Chaudhuri, and D. S. Ray, J. Chem.
Phys. \textbf{112}, 8330 (2000); K.M. Rattray and A.J. McKane, J.
Phys. A \textbf{24}, 4375 (1991); \textit{Noise in Nonlinear
Dynamical Systems}, edited by F. Moss and P. V. E. McClintock
(Cambridge University Press, Cambridge, 1989), Vols.I-III; J.
Masoliver and J. M. Porr\`{a}, Phys. Rev. E \textbf{48}, 4309
(1993); S. J. B. Einchcomb and A. J. McKane, Phys. Rev. E \textbf{
49}, 259 (1994).

\bibitem{res} P. Resibois and M. dc Leener, \textit{Chemical Kinetic
Theory of Fluids} (Wiley-Interscience, NY, 1977).

\bibitem{her} E. Hershkovits and R. Hernandez, J. Chem. Phys.
\textbf{122}, 014509 (2005); H. W. Hsia, N. Fang and X. Lee, Phys.
Lett. A \textbf{215}, 326 (1996); A. N. Drozdov and S. C. Tucker, J.
Phys. Chem. B \textbf{105}, 6675 (2001).

\bibitem{moi} J.M. Moix and R. Hernandez, J. Chem. Phys. \textbf{122}, 114111
(2005).

\bibitem{pol} E. Pollak and A. M. Berezhkovskii, J. Chem. Phys. \textbf{
99}, 1344 (1993).

\bibitem{mar} F. Marchesoni, Chem. Phys. Lett. \textbf{110}, 20 (1984);
A. V. Barzykin and K. Seki, Europhys. Lett. \textbf{40}, 117 (1997);
Q. Long, L. Cao, Da-jin Wu and Zai-guang Li, Phys. Lett. A \textbf{
231}, 339 (1997); O. V. Gerashchenko, S. L. Ginzburg and M. A.
Pustovoit, JETP Lett. \textbf{67}, 997 (1998); Y. M. Blanter and M.
B\"{u}ttiker, Phys. Rev. Lett. \textbf{81}, 4040 (1998); R.
Krishnan, M. C. Mahato and A. M. Jayannavar, Phys. Rev. E \textbf{
70}, 021102 (2004).

\bibitem{har} K. Hara, N. Ito and O. Kajimoto, J. Chem. Phys. \textbf{
110}, 1662 (1999).

\bibitem{zwa} R. Zwanzig, J. Stat. Phys. \textbf{9}, 215 (1973); K.
Lindenberg and V. Seshadri, Physcia A \textbf{109}, 483 (1981); M.
I. Dykman and M. A. Krivoglaz, Phys. Status Solidi B \textbf{48},
497 (1971).

\bibitem{bra} J. Mencia Bravo, R. M. Velasco and J. M. Sancho, J. Math.
Phys. \textbf{30}, 2023 (1989).

\bibitem{dipole} L.D. Landau and E.M. Lifshitz,
\textit{The Classical Theory of Fields} (Pergamon, Oxford, 1975).

\bibitem{van1} N. G. Van Kampen, Phys. Rep. \textbf{24}, 171 (1976).

\bibitem{grote-hynes} R. F. Grote and J. T. Hynes,
J. Chem. Phys. \textbf{73}, 2715 (1980).

\bibitem{hanggi-mojtabai} P. H\"{a}nggi and F. Mojtabai,
Phys. Rev. A  \textbf{26}, 1168 (1982); P. H\"anggi, J. Stat. Phys.
\textbf{30}, 401 (1983).

\bibitem{carmeli-nitzan} B. Carmeli and A. Nitzan,
J. Chem. Phys. \textbf{79}, 393 (1983); Phys. Rev. A \textbf{29},
1481 (1984).

\bibitem{borkovec} J. E. Straub, M. Borkovec, and B. J. Berne,
J. Chem. Phys. \textbf{84}, 1788 (1986).

\bibitem{adelman} S. A. Adelman, J. Chem. Phys. \textbf{64}, 124 (1976).

\bibitem{mah} C. Mahanta and T. G. Venkatesh, Phys. Rev. E \textbf{58}, 4141
(1998); J. M. Sancho, A. H. Romero, and K. Lindenberg, J. Chem.
Phys. \textbf{109}, 9888 (1998); D. Barik, B. C. Bag and D. S. Ray,
J. Chem. Phys. \textbf{119}, 12973 (2003).

\bibitem{ra} C. R. Doering and J. C. Gadoua,
Phys. Rev. Lett. \textbf{69}, 2318 (1992); M. Marchi, F. Marchesoni,
L. Gammaitoni, E. Menichella-Saetta, and S. Santucci, Phys. Rev. E
\textbf{54}, 3479 (1996); M. Bogu\~{n}\'{a}, J. M. Porr\'{a}, J.
Masoliver, and K. Lindenberg, Phys. Rev. E \textbf{57}, 3990 (1998);
P. K. Ghosh, D. Barik, B. C. Bag and D. S. Ray, J. Chem. Phys.
\textbf{123}, 224104 (2005).

\bibitem{ast} R. D. Astumian and P. H\"{a}nggi, Phys. Today \textbf{55}, 33
(2002).

\bibitem{den}  S. I. Denisov, A. N. Vitrenko, W. Horsthemke, P.
H\"{a}nggi, Phys. Rev. E \textbf{73}, 036120 (2006).

\end{thebibliography}
\end{document}